\documentclass[runningheads]{llncs}

\usepackage{graphicx}

\begin{document}

\title{Cognitive Workload Associated with Different Conceptual Modeling Approaches in Information Systems}
\titlerunning{Measuring Cognitive Workload Associated with Conceptual Models}
%
\author{Andreas Knoben\inst{1}\orcidID{0000-0001-7636-3244} \and
Maryam Alimardani\inst{1}\orcidID{0000-0003-3077-7657} \and
Arash Saghafi\inst{2}\orcidID{0000-0002-5906-8400} \and
Amin K. Amiri\inst{2}\orcidID{0000-0001-5667-9620}}
\authorrunning{A. Knoben et al.}
%
\institute{Department of Cognitive Science and AI, Tilburg University, Tilburg 5037 AB, Netherlands \and
Department of Management, Tilburg University, Tilburg 5037 AB, Netherlands
\email{a.j.knoben@tilburguniversity.edu}}
\maketitle              
\begin{abstract}
Conceptual models visually represent entities and relationships between them in an information system. Effective conceptual models should be simple while communicating sufficient information. This trade-off between model complexity and clarity is crucial to prevent failure of information system development. Past studies have found that more expressive models lead to higher performance on tasks measuring a user's deep understanding of the model and attributed this to lower experience of cognitive workload associated with these models. This study  examined this hypothesis by measuring users' EEG brain activity while they completed a task with different conceptual models. 30 participants were divided into two groups: One group used a low ontologically expressive model (LOEM), and the other group used a high ontologically expressive model (HOEM). Cognitive workload during the task was quantified using EEG Engagement Index, which is a ratio of brain activity power in beta as opposed to the sum of alpha and theta frequency bands. No significant difference in cognitive workload was found between the LOEM and HOEM groups indicating equal amounts of cognitive processing required for understanding of both models. The main contribution of this study is the introduction of neurophysiological measures as an objective quantification of cognitive workload in the field of conceptual modeling and information systems.

\end{abstract}

\section{Introduction}\label{sec:introduction}
Conceptual models are formal representations of the real world that are used for the purposes of understanding and communication between stakeholders, analysts, and developers of information systems \cite{mylopoulos1992conceptual,wand2002research}. They are used in the development of information systems with the aim to communicate the requirements of an application domain in such a way that it guides effective design of the system \cite{gemino2005complexity}. A widely used type of conceptual model is entity-relationship diagrams (ERDs), which communicate entities, i.e., constructs a business needs to remember in order to run the business, as well as relationships between those entities.

The use of ontology, a ``branch of philosophy that deals with the order and structure of reality'' \cite{angeles1981dictionary}, has been proposed to guide the creation of effective conceptual models \cite{wand1993ontological}. To reduce the chance of failure in the development process of an information system, a conceptual model must faithfully represent the relevant aspects of the modeled domain, and it must be clear and understandable to ``users'' of the model \cite{bera2014research}. Therefore, it has to be simple while communicating sufficient information to the user. Conceptual models that represent reality more closely are considered more ontologically expressive, but this is achieved possibly at the cost of increased complexity of the model \cite{wand1993ontological}.

The trade-off between ``expressiveness'' and “simplicity” has been studied in a number of previous papers and addressed in the meta-analysis by Saghafi and Wand (2020) \cite{saghafi2020meta}. It was found that ERDs with higher ontological expressiveness significantly improved users’ understanding of an application domain across different conceptual modeling grammars \cite{saghafi2020meta} as well as their performance on tasks that required a deeper understanding of the model, e.g., problem solving \cite{burton2008effects,gemino2005complexity}. These findings indicated a positive influence of ontological expressiveness on user interaction with the model allowing for better performance using the system. Gemino \& Wand (2005) discussed this finding from a cognitive load theory perspective, arguing that models with higher ontological expressiveness require less cognitive workload from the user as they represent the domain more clearly and the user is better able to understand and retrieve information \cite{gemino2005complexity}. However, they never provided any objective quantification of cognitive workload to support this argument.

A more direct approach to assess users' cognitive effort when they interact with and develop an understanding of a conceptual model is to measure brain responses during the task using electroencephalography (EEG). Past neuroscientific studies indicate that changes in cognitive workload are reflected in frequency band powers of EEG signals \cite{alimardani2021assessment,berka2007eeg,khedher2019tracking,pope1995biocybernetic}. A neural metric that is often employed to evaluate workload is the “EEG Engagement Index”, which is defined as the ratio between beta power and the sum of theta and alpha powers extracted from an EEG signal \cite{alimardani2021assessment,berka2007eeg}. This index has already been employed and verified in multiple human-technology interaction studies, however its application for user interaction with information systems remains unexplored.   

This study, for the first time, investigates neurophysiological metrics of cognitive workload when two groups of users work with conceptual models of different ontological expressiveness as suggested by Gemino \& Wand (2005) \cite{gemino2005complexity}. The research question is therefore formulated as below:

\textit{Is there a difference in cognitive workload as measured by brain activity when users retrieve information from two models with a different level of ontological expressiveness?}

\section{Methods}\label{sec:methods}
\subsection{Participants}
Thirty university students with a mean age of 22.13 years ($SD = 4.81$) participated in this study. They were randomly assigned to one of the two groups: One group worked with a low ontologically expressive model (the LOEM group), and the other worked with a high ontologically expressive model (the HOEM group). The study was approved by the Research Ethics Committee of the Tilburg School of Humanities and Digital Sciences (REDC 2021.144). Prior to the experiment, participants read an information letter and signed an informed consent form.

\subsection{Experimental Task}
\begin{figure}[h]
    \centering
    \includegraphics[width=\textwidth]{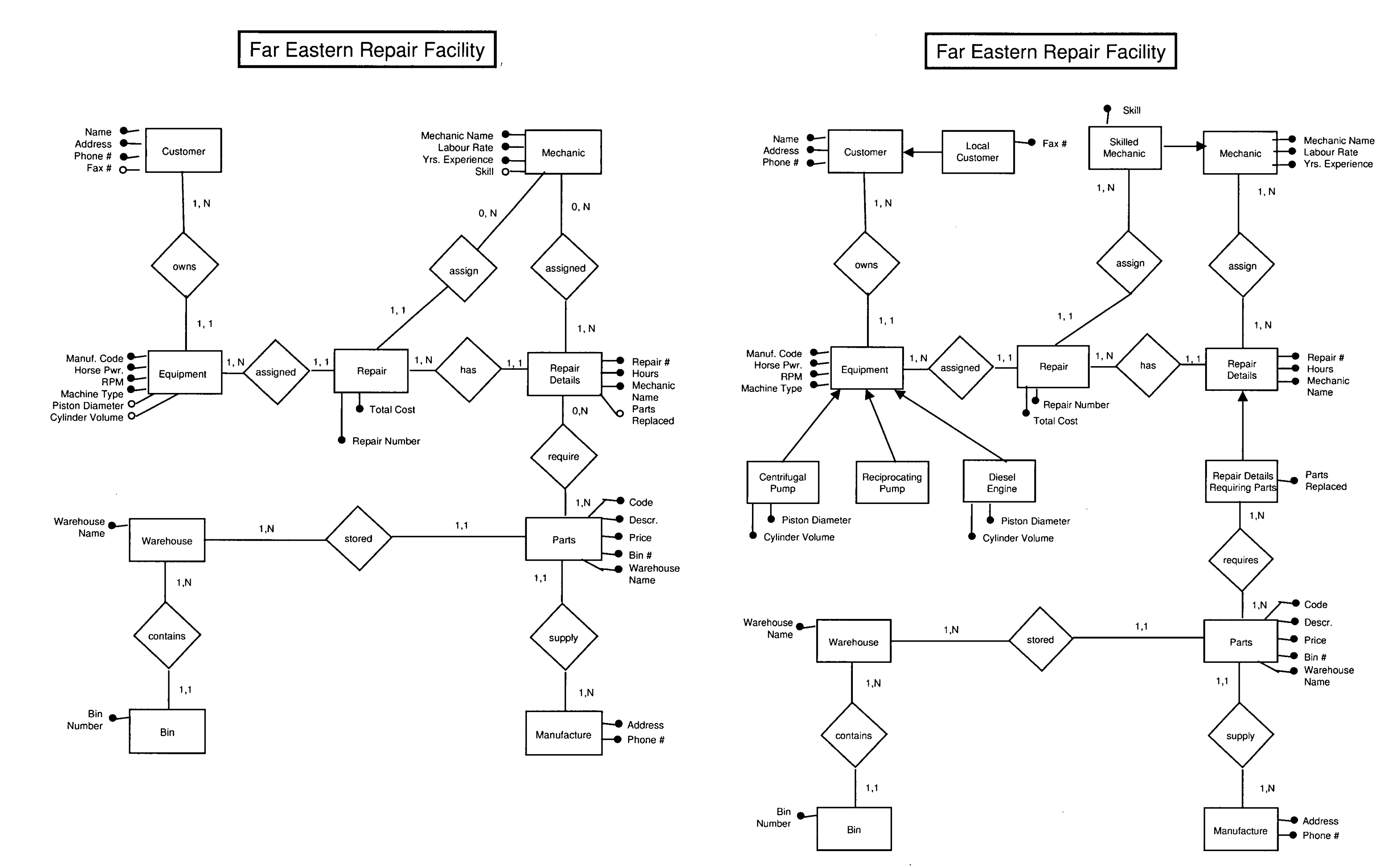}
    \caption{The low ontologically expressive (left) and high ontologically expressive (right) models used in this study, taken from \cite{gemino1999empirical}.}
    \label{fig:erds}
\end{figure}

Two conceptual models with different ontological expressiveness but presenting the same business domain were chosen for this study. The ERDs belonging to the selected domain, namely a machine repair facility, were taken from \cite{gemino1999empirical}, and can be viewed in Figure \ref{fig:erds}. The left model consists of entities with optional properties that lack certainty and hence is considered a low ontologically expressive model. The right model consists of mandatory properties that communicate unambiguous information and hence is considered as the high ontologically expressive model. Each participant was randomly assigned to one of these models and was instructed to complete missing information in a fill-in-the-blanks task using the ERD in front of them.

\subsection{EEG Recording}
EEG signals were recorded from 16 electrodes according to the 10–20 international system (Fp1, Fp2, F3, Fz, F4, T7, C3, Cz, C4, T8, P3, Pz, P4, PO7, PO8, and Oz). This selection allowed recording from all brain areas, i.e., left and right hemisphere and midline frontal, temporal, parietal, and occipital regions. The reference was set on the right earlobe and a ground electrode on AFz. A g.Nautilus amplifier (g.tec Medical Engineering, Austria) was used to collect the signals at the sampling rate of 250 Hz.

\subsection{Experiment Design}
Before the day of the experiment, the participants received an email with preparation instructions for the EEG, and a video tutorial that explained ERDs. This tutorial aimed to familiarize the participants with ERDs prior to the experiment. Upon arrival on the experiment day, first, the experimental procedure was explained to the participant and informed consent was obtained. Next, participants filled in a demographic questionnaire and the EEG electrodes were placed on the participant's head. When this was done, the participant sat behind a desk with two computer screens and began the experimental tasks.

Before the main task, EEG signals in a 3-min baseline were recorded, while participants looked at a fixation cross on the screen. Then, they read a text about the domain that the ERD in the task was going to model (machinery repair), and studied the diagram itself. After that, the main task started where they answered fill-in-the-blank questions by retrieving information from the model, which was presented on a second screen. The diagram remained visible on the screen, so that the participant could refer to it while working on the task. The mean time it took for the participants to complete the task was 8.78 minutes ($SD = 2.06$) for the LOEM group and 10.51 minutes ($SD = 3.26$) for the HOEM group.  Once the experiment was finished, the EEG cap was removed and the participant was briefed and thanked for their participation.

\subsection{Analysis}
The raw EEG data was pre-processed using MATLAB R2021b (Mathworks Inc., MA) and the EEGLAB toolbox \cite{delorme2004eeglab}. The pre-processing steps included discarding bad channels if necessary, applying a band pass filter of 0.5 to 60 Hz, rejecting artifacts in the data temporally, and rejecting eye artifact components using Independent Component Analysis. Then, spectral analysis was conducted on the clean data using the \verb|spectopo| function from the EEGLAB toolbox \cite{delorme2004eeglab} and the mean beta (13-30 Hz), alpha (8-13 Hz), and theta (4-8 Hz) powers for each channel and participant were extracted. 

Using these powers, the EEG Engagement Index was calculated according to the following equation \cite{pope1995biocybernetic}:

\begin{equation}\label{eq:eeg-ei}
    E = \frac{\beta}{\alpha + \theta}
\end{equation}

\noindent where $E$ is the EEG Engagement Index and $\beta$, $\alpha$, and $\theta$ are the mean power in the beta, alpha, and theta frequency bands, respectively \cite{alimardani2021assessment,chiang2018eeg,pope1995biocybernetic}. For every participant, the EEG Engagement Index was baseline-corrected, i.e., the baseline EEG Engagement Index was subtracted from the task EEG Engagement Index. 

The baseline-corrected EEG Engagement Indices were compared between the LOEM and HOEM groups in R (version 4.1.2). For the data of each channel, it was first determined whether it was normally distributed using a Shapiro-Wilk test. If the data was normally distributed, a Welch’s independent $t$-test was used to compare the groups. If the normality assumption was not met, a Mann-Whitney U test was used. Because the groups were compared 16 times, the significance level was adjusted to $0.05/16 = 0.003125$ using the Bonferroni method.

\section{Results}
\begin{figure}[h!]
    \centering
    \includegraphics[width=0.95\textwidth]{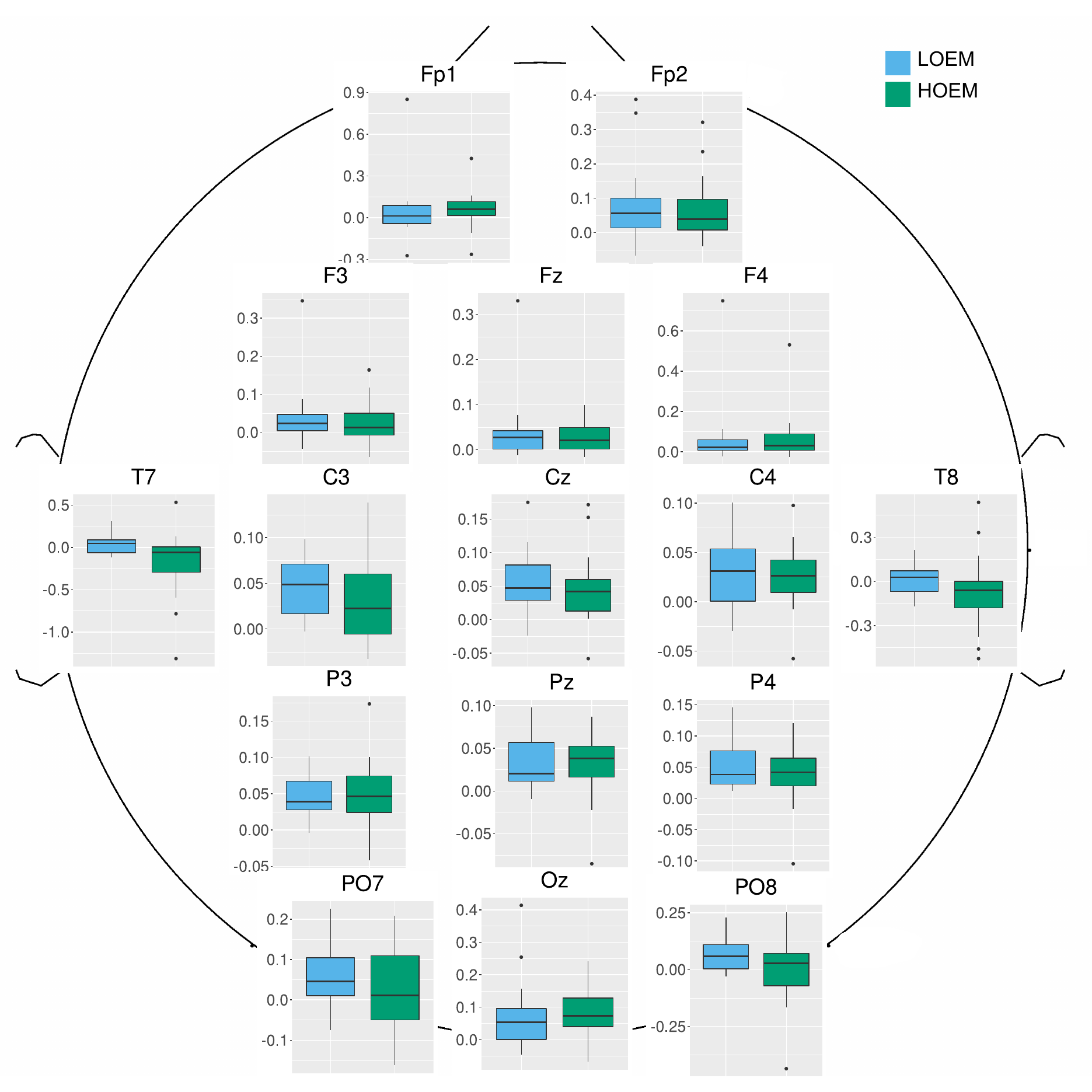}
    \caption{Plots showing the EEG Engagement Index obtained from participants working with low ontologically expressive (LOEM group) and high ontologically expressive models (HOEM group) at all electrode locations.}
    \label{fig:boxplots}
\end{figure}

The EEG Engagement Indices for each channel are visualized in Figure \ref{fig:boxplots}. Additionally, the results of the statistical test in each EEG channel is reported in Table \ref{tab:results}. No significant differences were observed in EEG-based cognitive workload that was produced by the subjects in LOEM and HOEM groups.

\begin{table}
    \caption{Summary of statistical analysis comparing EEG Engagement Index between LOEM and HOEM groups at each channel. The outcome of the Welch’s independent $t$-test (left) and Mann-Whitney U test (right) are separated.}
    \label{tab:results}
    \centering
    \begin{tabular}[t]{|l|l|l|l|}
        \hline
        Channel & $t$ & $df$ & $p$ \\
        \hline
        T7  & 1.97 & 16.03 & .07 \\
        C3  &  .73 & 23.78 & .47 \\
        Cz  &  .49 & 27.44 & .63 \\
        C4  &  .39 & 28.00 & .70 \\
        T8  & 1.02 & 18.59 & .32 \\
        P3  & -.20 & 23.31 & .84 \\
        Pz  &  .27 & 25.25 & .79 \\
        P4  &  .87 & 25.74 & .39 \\
        PO7 &  .83 & 24.37 & .42 \\
        PO8 & 1.60 & 19.68 & .13 \\
        \hline
    \end{tabular}
    \quad
    \begin{tabular}[t]{|l|l|l|}
        \hline
        Channel & $W$ & $p$ \\
        \hline
        Fp1 &  70 & .33 \\
        Fp2 & 118 & .84 \\
        F3  & 124 & .65 \\
        Fz  & 107 & .84 \\
        F4  & 106 & .81 \\
        Oz  &  54 & .79 \\
        \hline
    \end{tabular}
\end{table}

\section{Discussion}
In this study, EEG signals were employed to obtain a neurophysiological and an objective measure of cognitive workload when two groups of participants used conceptual models with different ontological expressiveness (low versus high) to conduct an information retrieval task. Cognitive workload was quantified by the EEG Engagement Index (Eq. \ref{eq:eeg-ei}) obtained from 16 channels distributed over the scalp. Our results showed no significant difference in EEG Engagement Index between the two groups at any of the investigated brain regions.

The findings presented here are inconsistent with the hypothesis of Gemino \& Wand (2005) that a high ontologically expressive model would require less workload \cite{gemino2005complexity}. They base their hypothesis on the argument that mandatory properties lead to a lower cognitive workload compared to optional properties, because models with mandatory properties have a clearer structure, whereas models with optional properties require extra reasoning in order to understand the domain \cite{gemino2005complexity}. However, our results, providing an objective quantification of cognitive workload from neural activity of participants, did not support such a hypothesis, suggesting that both groups recruited similar amounts of cognitive resources  when processing optional and mandatory properties of the models.

A reason why model comparison in this study did not show a significant difference may be due to the relatively small sample size where only 15 subjects were present in each group. Particularly with individual differences present in EEG signals, future research should expand the study with a larger sample to  be able to provide a reliable comparison. 

Another reason could be participants' lack of experience in working with ERD models in this study. In the experiment of Gemino \& Wand, participants were business students who had taken at least one course related to Management Information Systems and were thus familiar with conceptual ERD models \cite{gemino2005complexity}. Participants in our study were novice to this topic, and although they watched an introductory video about ERD before the task, their previous experience with conceptual models was almost non-existent. It could be speculated that such difference in background knowledge between the participants of the two studies led to different outcomes.

Finally, in this study, EEG Engagement Index was only compared during one task which required filling in missing information, whereas Gemino \& Wand (2005) proposed multiple tasks in their study among which the problem-solving task was argued to illustrate deeper understanding of a model \cite{gemino2005complexity}. Future research should provide a more comprehensive analysis of brain activity as well as performance on the task when users interact with ERD models from various domains and conduct multiple tasks that target different levels of cognitive functioning. 
 
In sum, the current study provides preliminary insights for future employment of neurophysiological measures in evaluation of conceptual modeling methods. Using a quantified indicator of cognitive workload borrowed from neuroergonomics research,  we propose that the trade-off between clarity and complexity in conceptual models can be addressed by measuring brain activity to strike an optimal balance between these two criteria from the user’s perspective \cite{burton2008effects}. 

\section{Conclusion}
This study aimed to quantitatively assess user interaction with conceptual models in information systems using EEG brain activity. Cognitive workload was extracted from EEG signals when two groups of users worked with a low ontologically expressive model versus a high ontologically expressive model. Contrary to the predictions of previous research, no significant difference was found in cognitive workload as measured by brain activity in any of the EEG electrodes. This study demonstrates preliminary results for employment of neurophysiological measurements in evaluation of Management Information Systems. Future work can validate the precision and reliability of this method by employing larger groups of participants and including more complex tasks (e.g., problem-solving) to examine deeper levels of model understanding in brain responses. 

%
%
%

\bibliographystyle{splncs04}
\bibliography{references}

\end{document}